\documentclass{article}

\usepackage[final,nonatbib]{nips_2018}

\usepackage[utf8]{inputenc} 
\usepackage[T1]{fontenc}    
\usepackage{hyperref}       
\usepackage{url}            
\usepackage{booktabs}       
\usepackage{amsfonts}       
\usepackage{nicefrac}       
\usepackage{microtype}      

\usepackage{amsmath}
\usepackage{amsfonts}
\usepackage{amssymb}
\usepackage{graphicx}%
\usepackage{booktabs}

\newcommand{\cond}{\,|\,}

\DeclareMathOperator{\Gam}{{Gamma}}

\DeclareMathOperator{\Poi}{{Poisson}}
\DeclareMathOperator{\Dir}{{Dir}}

\DeclareMathOperator{\Unif}{\mathcal{U}}

\newcommand{\Bz}{\mathbf{z}}

\newcommand{\Bomega}{\boldsymbol{\omega}}

\newcommand{\Btheta}{\boldsymbol{\theta}}

\renewcommand{\epsilon}{\varepsilon}

\usepackage{dsfont}

\newcommand{\stA}{ST$\,$A, str$\,$X $\rightarrow$ ST$\,$A, str$\,$X}
\newcommand{\stB}{ST$\,$A, str$\,$X $\rightarrow$ ST$\,$A, str$\,$Y}
\newcommand{\stC}{ST$\,$A $\rightarrow$ ST$\,$B}
\newcommand{\stD}{ST$\,$A, str$\,$X $\rightarrow$ $\emptyset$}
\newcommand{\stE}{$\emptyset$ $\rightarrow$ ST$\,$A, str$\,$X}

\allowdisplaybreaks 

\newcommand{\neff}{n_{\text{eff}}}
\newcommand{\ddist}{D_{0}}
\newcommand{\dclear}{D}
\newcommand{\dist}{\text{dist}}

\usepackage{wrapfig}

\usepackage{authblk}

\title{A Bayesian model of acquisition and clearance of bacterial colonization}

\author{Marko J\"{a}rvenp\"{a}\"{a}$^{1,2}$, Mohamad R. Abdul Sater$^2$, Georgia K. Lagoudas$^{3,4}$, Paul C. Blainey$^{3,4}$, Loren G. Miller$^5$, James A. McKinnell$^{5,6}$, Susan S. Huang$^7$, Yonatan H. Grad$^{2,*}$, Pekka Marttinen$^{1,*}$ \\ \vspace{-0.3cm} $^1$Aalto University 
$^2$Harvard TH Chan School of Public Health 
$^3$MIT
$^4$Broad Institute, MIT and Harvard
$^5$LA Biomedical Research Institute, Harbor-UCLA Medical Center
$^6$LA County \\ Department of Public Health, Acute Communicable Disease Control Unit
$^7$University of California, Irvine School of Medicine
(*equal contribution)}

\begin{document}

\maketitle

\begin{abstract}
  Bacterial populations that colonize a host play important roles in host health, including serving as a reservoir that transmits to other hosts and from which invasive strains emerge, thus emphasizing the importance of understanding rates of acquisition and clearance of colonizing populations. Studies of colonization dynamics have been based on assessment of whether serial samples represent a single population or distinct colonization events. A common solution to estimate acquisition and clearance rates is to use a fixed genetic distance threshold. However, this approach is often inadequate to account for the diversity of the underlying within-host evolving population, the time intervals between consecutive measurements, and the uncertainty in the estimated acquisition and clearance rates. Here, we summarize recently submitted work \cite{jarvenpaa2018named} and present a Bayesian model that provides probabilities of whether two strains should be considered the same, allowing to determine bacterial clearance and acquisition from genomes sampled over time. We explicitly model the within-host variation using population genetic simulation, and the inference is done by combining information from multiple data sources by using a combination of Approximate Bayesian Computation (ABC) and Markov Chain Monte Carlo (MCMC). We use the method to analyse
a collection of methicillin resistant Staphylococcus aureus (MRSA) isolates. 
\end{abstract}

\section{Introduction}

Colonizing bacterial populations are often the source of infecting strains and transmission to new hosts \cite{young2012evolutionary,young2017severe,gordon2017whole,alam2015transmission,coll2017longitudinal}, making it important to understand the dynamics of these populations and the factors that contribute to persistent colonization and to the success or failure of clinical decolonization protocols. The study of colonization dynamics is based on inferring whether bacteria from samples collected over time represent the same population or distinct colonization events, thereby permitting calculation of rates of acquisition and clearance \cite{calderwood2015editorial,miller2014dynamics}. Whole genome sequencing has provided a detailed measure of genetic distance between isolates, which can then be used to infer the relationship between them \cite{didelot2016within,worby2014within,price2017transmission,uhlemann2014molecular}. While to date most studies have used genetic distance thresholds as the basis for determining the relationship between isolates \cite{price2017transmission,didelot2016within}, in this text we improve on these heuristic strategies and present a robust and accurate fully probabilistic model that provides probabilities of whether two strains should be considered the same. 

The Bayesian statistical framework allows to combine information from multiple data sources. In this approach, a prior distribution is updated using the laws of probability into a posterior distribution in the light of the observations, and this can be repeated multiple times with different data sets \cite{Gelman2013,Ohagan2004}. Approximate Bayesian computation (ABC) is particularly useful with population genetic models, where the likelihood function may be difficult to specify explicitly, but simulating the model is feasible \cite{Beaumont2002,Lintusaari2016}. ABC has recently been introduced in bacterial population genetics \cite{ansari2014inference,numminen2013estimating,jarvenpaa2016gaussian,de2017bacterial}. Here, we present a Bayesian model for determining whether two genomes should be considered the same strain, enabling a strategy grounded in population genetics to make inferences about acquisition and clearance from data of closely related genomes. Benefits of this approach include: rigorous quantification of uncertainty, explicit statement of modeling assumptions (open for criticism and further development), and straightforward utilization of multiple data sources. We demonstrate these benefits by analyzing a large collection longitudinally collected MRSA genomes, obtained through a clinical trial (Project CLEAR) to evaluate the effectiveness of an MRSA decolonization protocol \cite{clear}. 

\section{Data sources}
\label{sec:data}

One input data item from \cite{clear} for our model consists of a pair of genomes that are of the same sequence type (ST), sampled from the same individual at two consecutive time points (or possibly with an intervening time point with no samples or a sample of a different ST). All the sampled genomes are from nasal swabs. Each of these pairs of consecutive genomes is summarized in terms of two quantities: the distance between the genomes $d_{i}\in\{0,1,2,\ldots\}$ and the difference between their sampling times $t_{i}>0$ (see Fig~\ref{fig:data_example}). Hence, the observed data, which we denote by $\dclear$, can be written as consisting of pairs $(d_{i},t_{i}),$ $i=1,\ldots,N$.

As external data we use measurements from eight patients colonised with MSSA \cite{gordon2017whole}, comprising nasal swabs from two time points for each patient, such that the acquisition is known to have happened approximately just before the first swab. Multiple genomes were sequenced from each sample, and the distributions of pairwise distances between the genomes provide snapshots to the within-host variability at the two time points for each individual, and these distance distributions are used as additional data. The data set also contains observations from an additional 13 patients from \cite{golubchik2013within}, denoted by letters from A to M in \cite{gordon2017whole}. For these patients, distance distributions from only one time point are available. The data from the $8+13$ patients is jointly denoted by $\ddist$.

\section{Outline of the model and the inference algorithms}
\label{sec:model}

\begin{figure}[ht]
\begin{center}
\includegraphics[width=0.76\textwidth]{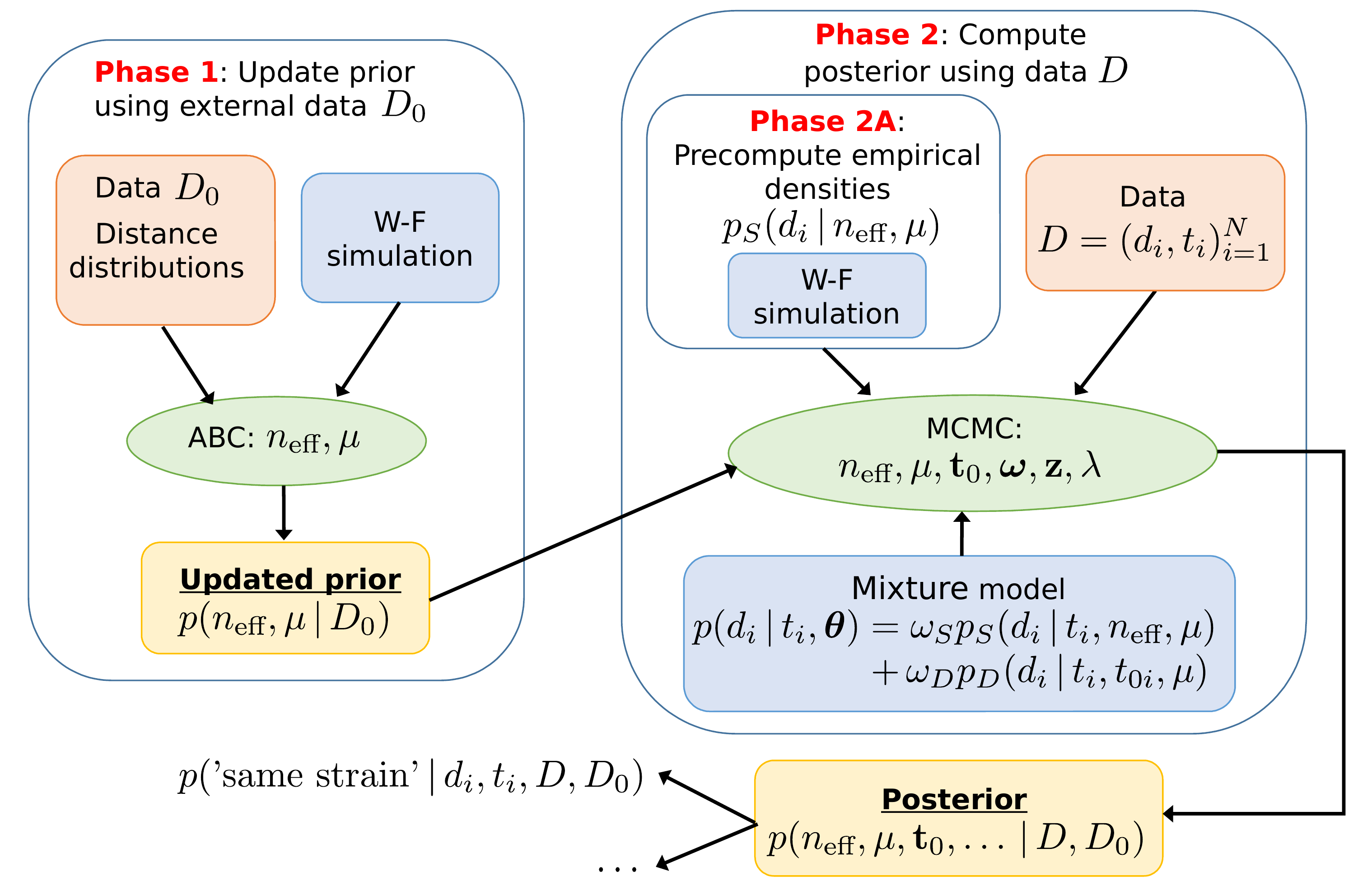}
\caption{{Overview of the modeling and data fitting steps.} 
} \label{fig:overview}
\end{center}
\end{figure}

Overview of the proposed approach, including data sets, models, and methods for inference, is outlined in Fig~\ref{fig:overview} and discussed in the following in more detail. 
An essential part of our approach is a population genetic simulation which allows us to model the within-host variation, and hence make probabilistic statements of the plausibilities of the 'same strain' vs. 'different strain' cases. For this purpose, we adopt the common Wright-Fisher (W-F) simulation model, see e.g.~\cite{schierup2010coalescent}, with a constant mutation rate $\mu$ and effective population size $\neff$, which are estimated from the data. The simulation is started with all genomes being the same, which corresponds to a biological scenario according to which a colonization begins with a single isolate multiplying rapidly until reaching the maximum 'capacity', followed by slow diversification of the population. 

\begin{figure}[ptb]
\begin{center}
\includegraphics[width=0.84\textwidth]{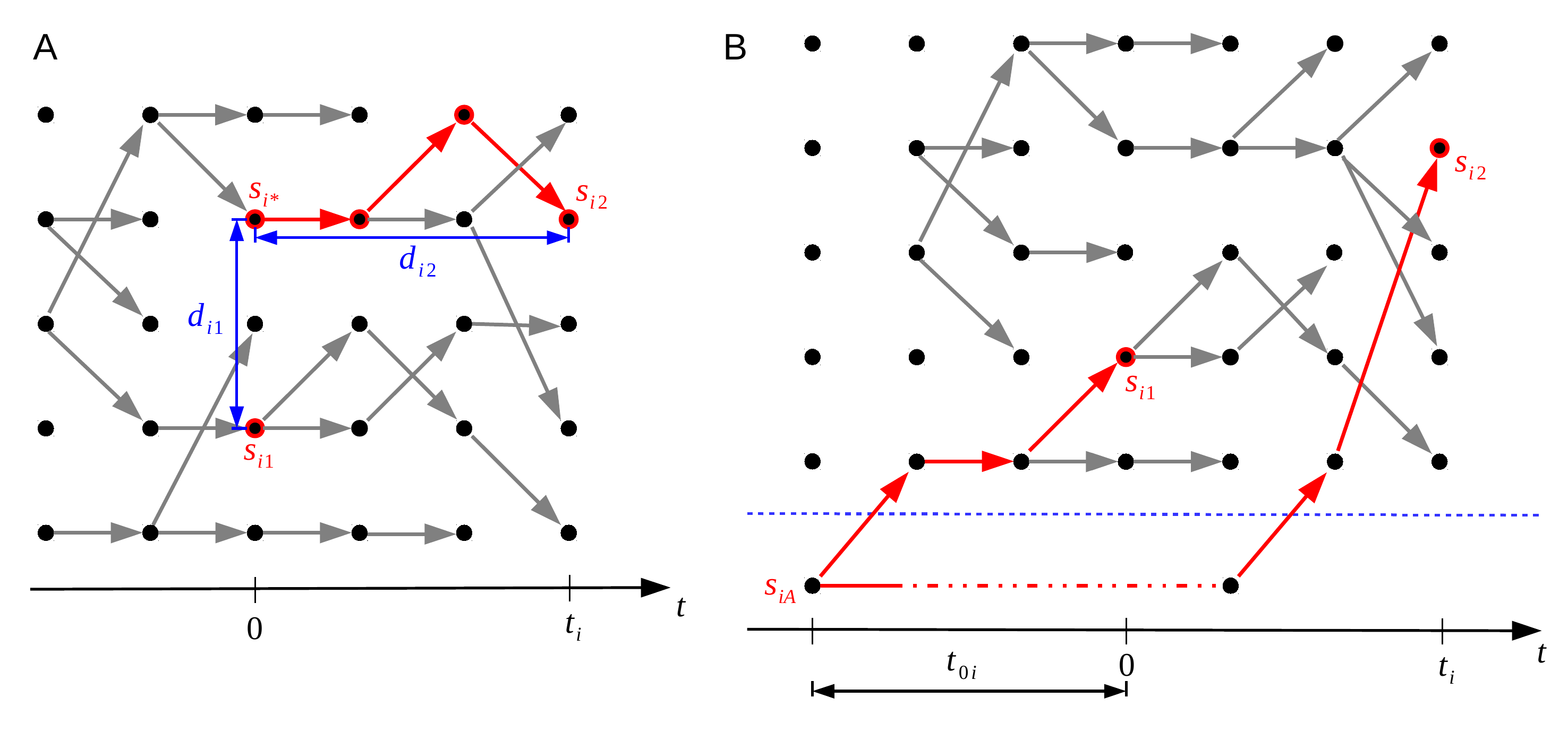}
\caption{{Outline of the 'same strain' and 'different strain' models.} Model $p_D$ on the panel A represents the situation where the genomes denoted by $s_{i1}$ and $s_{i2}$ are of the same strain. Model $p_S$ on the panel B shows the case where these genomes are of different strains. 
} \label{fig:strain_figs}
\end{center}
\end{figure}

Let $(s_{i1},s_{i2})$ denote a pair of genomes with distance $d_i$, sampled from a patient at two consecutive time points with time difference $t_i$. Here we present a model, i.e., a probability distribution $p_S(d_i \cond t_i,\neff,\mu)$, which tells what kind of distances we should expect if the genomes are from the same strain. We model $d_i$ as $d_i = d_{i1} + d_{i2}$ where we have defined $d_{i1} = \dist(s_{i1}, s_{i*})$ and $d_{i2} = \dist(s_{i*}, s_{i2})$,
where $\dist(\cdot,\cdot)$ is a distance function that tells the number of mutations between its arguments, and $s_{i*}$ is the unique ancestor of $s_{i2}$ that was present in the host when $s_{i1}$ was sampled, and which has descended within the host from the same genome as $s_{i1}$ (see Fig~\ref{fig:strain_figs}A). 
The probability distribution of $d_{i1}$ which is denoted by $p_{\text{sim}}(d_{i1}\cond\neff,\mu)$, and which is not available analytically and does not depend on $t_i$, represents the within-host variation at a single time point, and we define it implicitly as 
\begin{equation}\label{eq:first_dist}
    p_{\text{sim}}(d_{i1} \cond \mu,\neff) = \text{WF-simulator}(d_{i1} \cond \mu, \neff).
\end{equation}
The distribution of $d_{i2}$ is assumed to be
%
    $d_{i2} \cond \mu,t_{i} \sim \Poi(d_{i2}\cond\mu t_{i})$,
%
that is, mutations are assumed to occur according to a Poisson process with the rate parameter $\mu$.

Model $p_D$ represents the case that the genomes $s_{i1}$ and $s_{i2}$ are from different strains, which we define to mean that their most recent common ancestor (MRCA), denoted by $s_{iA}$, resided outside the host. The time between $s_{iA}$ and $s_{i1}$ is denoted by $t_{0i}$ (see Fig~\ref{fig:strain_figs}B). Under model $p_D$, we assume that the distribution of the distance $d_{i}$ is
\begin{equation}
p_D(d_{i} \cond \mu,t_{i},t_{0i})=\Poi(d_{i} \cond \mu(2t_{0i}+t_{i})),
\label{eq:model2_eq}
\end{equation}
where the values of $t_{0i}$ are unknown and will be estimated. 

With the two alternative models for the distance, we write the full model, which assumes that each distance observation is distributed according to
\begin{equation}
p\left( d_{i}\cond t_{i},\Btheta \right) = \omega_S p_{S}(d_{i}\cond t_{i},\neff,\mu) + \omega_D p_{D}(d_{i}\cond t_{i},t_{0i},\mu), \quad i=1,\ldots,N,
\label{eq:mixture_model}
\end{equation}
where $\Btheta$ denotes jointly all the parameters of the models, i.e., $\Btheta = (\neff,\mu,\omega_S,\omega_D,t_{01},\ldots,t_{0N})$. The parameter $\omega_S$ represents the proportion of pairs from the same strain and $\omega_D$ is the proportion of pairs from different strains, such that $\omega_S+\omega_D=1$.

Because the values of $t_{0i}$ in Eq~\ref{eq:model2_eq}, denoting the times to the MRCAs in case the sequences are different strains, are unknown, we model them as random variables and use the hierarchical prior distribution
\begin{equation}
t_{0i} \cond k, \lambda \sim \Gam(k, \lambda), i = 1,\ldots,N \quad \text{and} \quad \lambda \sim \Gam(\alpha,\beta). \label{eq:ti0_prior}
\end{equation}
The parameter $\lambda$ is thus shared between different $t_{0i}$ which allows us to learn about its distribution. We set $k=5$, $\alpha=2.5$, and $\beta=1600$, which approximately correspond to the mean and standard deviation of $5800$ and $8400$ generations, respectively. This weakly informative prior reflects the notion that different strains diverged on average approximately a year ago, but with a large variance.
Furthermore, we set $\Bomega = (\omega_S,\omega_D) \sim \Dir(1,1)$. 
%
As a prior for $(\neff,\mu)$, we use $\neff \sim \Unif(\{20,21,\ldots,10000\}), \mu \sim \Unif([a_{\mu},b_{\mu}])$ with $a_{\mu} = 0.00005$ and $b_{\mu} = 0.005$ mutations per genome per generation. We then use ABC inference, extensive W-F forward simulations and the external data $\ddist$ to obtain (approximate) posterior $p(\neff,\mu \cond \ddist)$ which is further used as a prior for the mixture model.

As in e.g.~\cite{Bishop2006}, we introduce hidden labels (denoted jointly by $\Bz$) which specify the component which generated each observation $d_{i}$. This allows to derive a Gibbs sampling algorithm for the posterior of the augmented model. To make the inference fast, we use various additional tricks, e.g.~we reparametrize the model and block some parameters to obtain better mixing of the MCMC. Since the densities in Eq.~\ref{eq:first_dist} are available only implicitly, we estimate them empirically with additional W-F simulations. Due to lack of space we omit further details of the resulting MCMC algorithm.


\section{Experiments and conclusion}
\label{sec:experiments}

The estimated 'global' parameters of the mixture model given the observed data sets $\dclear$ and $\ddist$ described in Section \ref{sec:data} are presented in Table~\ref{table1}.
We have also investigated the efficiency of the MCMC method using simulated data and some of the results are shown in the appendix. We computed acquisition and clearance rates using our model fitted to the data, and compared those to the ones obtained with the common strategy of using a fixed distance threshold of $40$ ('baseline' case). These results are shown in Table~\ref{table3} along with some statistics on the expected number of different events in the data.  We see that the threshold-based estimates are relatively similar to, and only slightly smaller than the estimates from our model. Importantly, while being consistent with the previous results, our model bypasses the
task of heuristically choosing a single threshold and adds uncertainty estimates around
the point estimates, crucial for drawing rigorous conclusions.

\begin{table}
\begin{minipage}{0.38\textwidth}
\centering
\caption{
{Posterior mean and 95\% credible interval (CI) for the 'global' parameters of the mixture model. 
The estimated 'nuisance' parameters $t_{0i}$ and class labels $\Bz_i$ are not shown here. 
}}
\begin{tabular}{lll}
\toprule
parameter & mean & 95\% CI \\
\midrule
$\neff$ & $1700$ & $[1300,2200]$ \\ 
$\mu$ ($\times 10^{4}$) & $7.6$ & $[6.0,9.2]$    \\ 
$\omega_S$ & $0.87$ & $[0.83,0.91]$    \\ 
$\omega_D$ & $0.13$ & $[0.09,0.17]$    \\ 
$\lambda$ ($\times 10^{5}$) & $7.3$ & $[5.8,9.0]$    \\ \bottomrule
\end{tabular}
\label{table1}
%
\end{minipage}
\hfill
\begin{minipage}{0.6\textwidth}
%
\centering
\small
\caption{
{Posterior means of different patterns of consecutive samples and the estimated acquisition and clearance rates (mean, 95\% CI in parenthesis, 'str' refers to strain).}}
\begin{tabular}{lll}
\toprule
event & expected number \\
\midrule
\stA & $231$ &\\
\stB & $34$ &\\
\stC & $45$ &\\
\stD & $104$ &\\
\stE & $21$ &\\
\toprule
rate parameter & post.~estimate & baseline \\
\midrule
acquisition rate $r_{\text{acq}}$ & $0.18 (0.17,0.19)$ & 0.16 \\
clearance rate $r_{\text{clear}}$ & $0.25 (0.24,0.25)$ & 0.24 \\
\bottomrule
\end{tabular}
\label{table3}
\end{minipage}
\end{table}

\begin{wrapfigure}{l}{0.35\textwidth}
\vspace{-0.5cm}
\begin{center}
\includegraphics[width=0.35\textwidth]{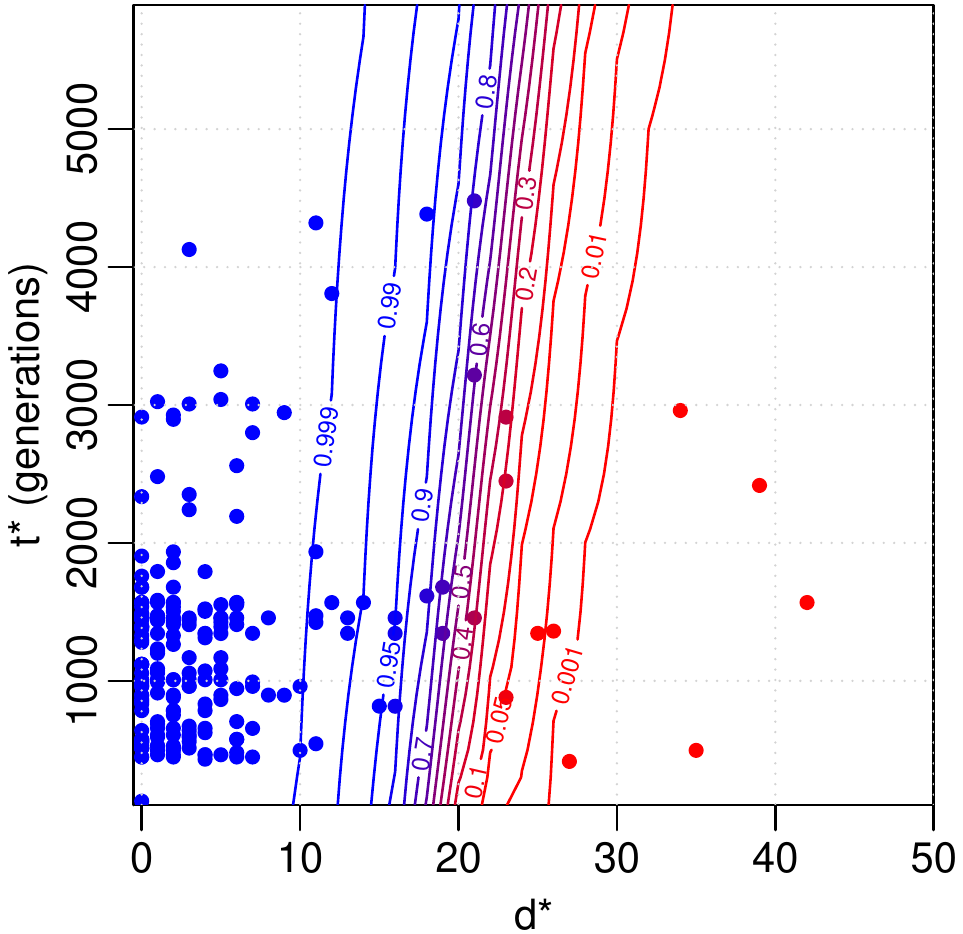}
\caption{Same strain probability of a new measurement with distance $d^*$ and time interval $t^*$. 
} \label{fig:pz}
\end{center}
\vspace{-1cm}
\end{wrapfigure}

Fig~\ref{fig:pz} shows the posterior distribution for the probability of the same strain case for a (hypothetical future) observation with distance $d^*$ and time difference $t^*$. Blue colour in the figure denotes high probability of the same strain. The corresponding $50\%$ classification curve is (almost) a straight line with a steep positive slope. This is as expected since the same strain model can explain a greater number of mutations when more time has passed. Approximately $20$ mutations draws the line between the same strain and different strains cases within the time difference up to $6000$ generations (approximately one year).

To summarize, we presented a model for the analysis of clearance and acquisition of bacterial colonization, which, unlike previous approaches, does not rely on a heuristic fixed distance threshold to determine whether genomes observed at different times points are from the same or different acquisition. Fully probabilistic, the model automatically provides uncertainty estimates for all relevant quantities and takes into account the variation in the time intervals between pairs of consecutive samples. As future work, we will extend the model to cover genomes sampled from multiple body sites.

%
\subsection*{Acknowledgments}

We acknowledge the computational resources provided by Aalto Science-IT project.




\bibliography{final_arxiv}
\bibliographystyle{abbrv}


\normalsize

\section*{Appendix}
\appendix

\subsection*{Visualisation of the MRSA data $\dclear$}

\begin{figure}[ht!]
\begin{center}
\includegraphics[width=0.75\textwidth]{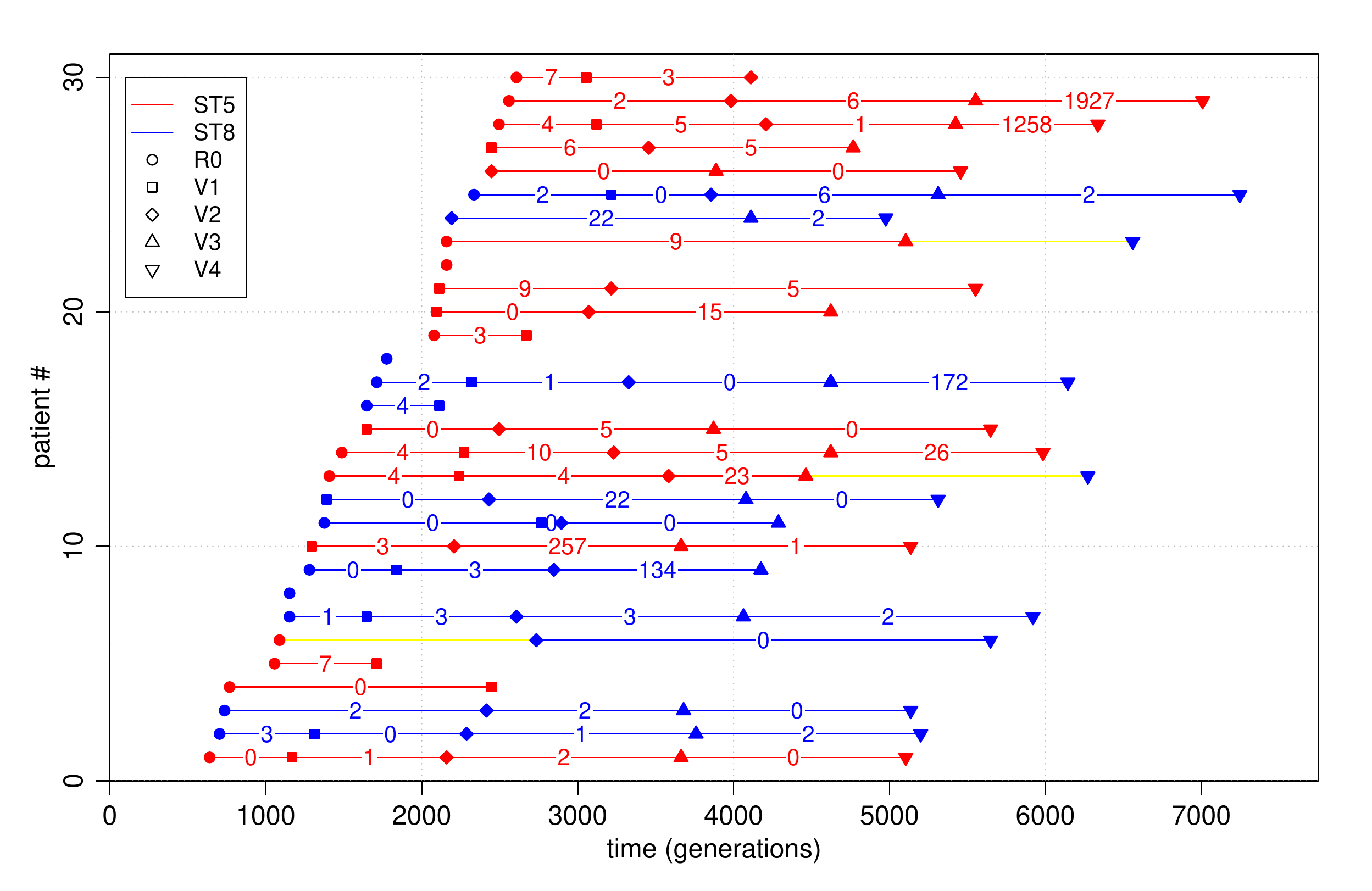}
\caption{{Illustration of a subset of the data used in the study.} Each row corresponds to one patient and only the first $30$ patients are shown. R0 is the initial hospital visit and V1, V2 etc.~are the further visits. Red colour refers to ST5 and blue to ST8 and the coloured numbers are the amount of mutations $d_i$. Yellow colour highlights the cases where the ST changes from ST 5 to ST 8. 
} \label{fig:data_example}
\end{center}
\end{figure}

An example of a typical individual-level longitudinally sampled data set (denoted by $\dclear$ in the main text) from a study population also used in our analysis is shown in Fig~\ref{fig:data_example}: each 'row' represents a patient, x-axis is time, and dots are the genomes sampled at multiple time points. Dot color refers to different, easily distinguishable, sequence types (ST). The coloured number between two consecutive samples reflects the distance between the genomes, and we see that even within the same ST the distances may vary considerably, and, therefore, determining whether the changes can be explained by within-host evolution only, is challenging. Intuitively, if two genomes are very similar, we interpret this as a single strain colonizing the host. On the other hand, two very different genomes, even if the same ST, are interpreted as two different strains, obtained either jointly or separately as two acquisitions.

\subsection*{Results with simulated data}

To empirically investigate the identifiability of the mixture model parameters and the correctness and consistency of our MCMC algorithm under the assumption that the model is specified correctly, we first fit the mixture model to simulated data. We generate artificial data from the mixture model with parameter values similar to the estimates for the observed data $\dclear$ from the next section. Specifically, we choose $\neff=2,137, \mu=0.0011, \omega_S=0.8, \lambda=0.0001$ and we repeat the analysis with various data sizes $N$. We use otherwise similar priors as for the real data except that, for simplicity, instead of using the prior obtained from the ABC inference, we use a uniform prior. We then fit the mixture model to the simulated data sets to investigate if the true parameters can be recovered (identifiability) and whether the posterior becomes concentrated around their true values when the amount of data increases (consistency).

\begin{figure}[ht!]
\begin{center}
\includegraphics[width=0.75\textwidth]{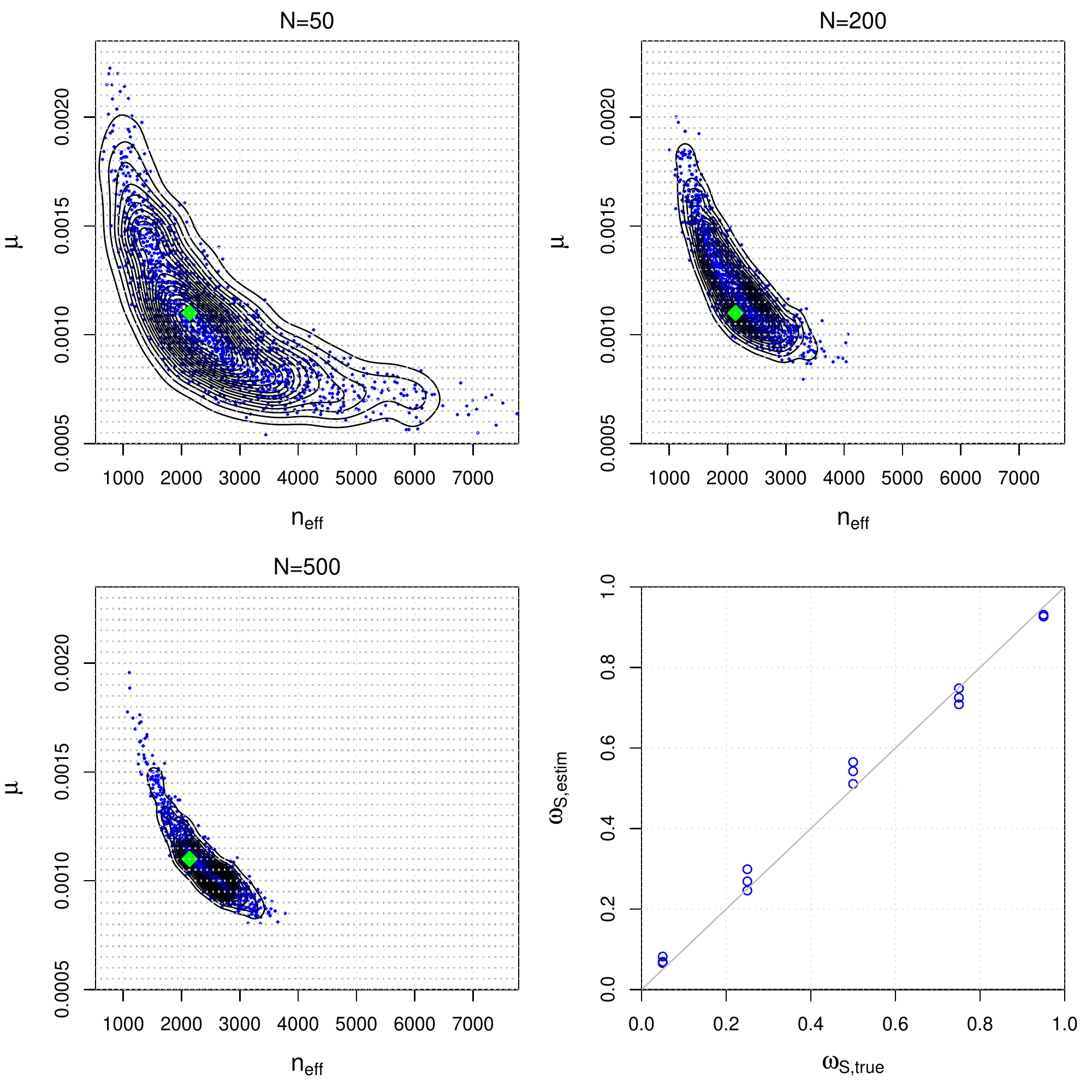}
\caption{{Illustration of the accuracy and consistency with synthetic data.} 
} \label{fig:fake_data_demo1}
\end{center}
\end{figure}

Results are illustrated in Fig~\ref{fig:fake_data_demo1}. The first three panels show the estimated posterior distributions for parameters $(\neff,\mu)$ of the mixture model using simulated data of different sizes $N$. The light grey dots denote the grid point locations needed for numerical computations. 

We see that the (marginal) posterior of $(\neff,\mu)$ is concentrated around the true parameter value that was used to generate the data (green diamond in the figure). 
Also, despite the fact that the number of parameters increases as a function of data size $N$ (because each data point $(d_i,t_i)$ has its own class indicator $\Bz_i$ and time to the most recent common ancestor $t_{0i}$ parameter), the marginal posterior distribution of $(\neff,\mu)$ can be identified and appears to converge to the true value as $N$ increases.

The panel in the lower right corner of Fig~\ref{fig:fake_data_demo1} shows results from an additional simulation experiment where the mixture model is fitted to data generated with different values for the $\omega_S$ parameter, which represents the proportion of pairs that are from the same strain. Other than that and the fact that we fixed $N=150$, the experimental design is the same as above. The results show that the estimated $\omega_S$ values generally agree well with the true values.
%

\end{document}